\documentclass[aps,prl,twocolumn,reprint,floatfix,longbibliography,superscriptaddress]{revtex4-2}
\usepackage[utf8]{inputenc} 
\usepackage{graphicx}
\usepackage{amsmath}
\usepackage{amsfonts}
\usepackage{mathtools}
\usepackage{float}
\usepackage{tabularx, booktabs}
\usepackage{color}
\usepackage{braket}
\usepackage{bm}
\usepackage{comment}

\usepackage{hyperref}

\hypersetup{
	colorlinks=true,
	linkcolor=blue,
	citecolor=blue,
	filecolor=magenta,
	urlcolor=cyan,
	breaklinks=true
}
\newcommand{\papertitle}{Dimer problem on a spherical surface}
\newcommand{\lptms}{Universit\'e Paris-Saclay, CNRS, LPTMS, 91405 Orsay, France}
\newcommand{\ICFO}{ICFO-Institut de Ciencies Fotoniques, The Barcelona Institute of Science and Technology, Castelldefels (Barcelona) 08860, Spain.}

\DeclareSymbolFont{sfletters}{OML}{cmbrm}{m}{it}

\begin{document}
	
\title{\papertitle}
\author{A. Tononi}
\affiliation{\ICFO}
\author{D. S. Petrov}
\affiliation{\lptms}
\author{M. Lewenstein}
\affiliation{\ICFO}
\affiliation{ICREA, Pg. Lluís Companys 23, 08010 Barcelona, Spain}

\date{\today}

\begin{abstract}
We solve the problem of a dimer moving on a spherical surface and find that its binding energy and wave function are sensitive to the total angular momentum. The dimer gets squeezed in the direction orthogonal to the center-of-mass motion and can qualitatively change its geometry from two-dimensional to one-dimensional. 
These results suggest that combining the curved geometry with finite angular momentum may give rise to qualitatively new many-body phenomena in ultracold shell-shaped gases.
\end{abstract}

\maketitle 
The problem of two interacting bodies has a central importance in diverse areas of physics ranging from celestial mechanics and general relativity \cite{einstein1938, barker1975, buonanno1999} to classical electrodynamics \cite{schild1963}.
In quantum mechanics, it underlies the solution of the hydrogen atom and the theory of scattering \cite{Landau}.
Two-body physics also rules the thermodynamic description of ultracold atomic gases~\cite{dalfovo1999,giorgini2008} since their interaction range is much smaller than their de Broglie wavelengths and average interparticle distances. 
In particular, the zero-range scattering problem has been solved in three-dimensional free space \cite{Landau}, in quasi-one-dimensional~\cite{olshanii1998} and in quasi-two-dimensional~\cite{petrov2001} {geometries, and} the spectrum of a harmonically-confined pair of atoms was obtained in any spatial dimension~\cite{busch1998}. 
These solutions are crucial for understanding Feshbach resonances in trapped gases \cite{stoferle2006, thalhammer2006}, the crossover from the Bardeen-Cooper-Schrieffer state to the Bose-Einstein condensate of molecules (BCS-BEC crossover) in fermionic mixtures \cite{zwerger2011}, two-dimensional universal thermodynamics \cite{hung2011, yefsah2011}, solitons and nonlinear states \cite{bakkalihassani2021, chen2021}, and many other phenomena in ultra-cold gases.

In the above cases, the solution of the two-body problem is simplified by its \textit{separability} into two independent single-particle problems: one for the center-of-mass free dynamics, another for the relative motion of the particles. 
The separability is however not assured if the particles are constrained to move in optical lattices \cite{fedichev2004,schneider2009}, in mixed dimensional setups \cite{massignan2006, nishida2008, lamporesi2010, xiao2019}, in anharmonic \cite{bolda2005,sala2016} or species-dependent harmonic \cite{peano2005,melezhik2009} potentials, 
or in spatial domains which are compact or curved.
In particular, solving the two-body problem in curved setups is more difficult than in flat counterparts, but fundamentally valuable for discovering new quantum mechanical behaviors induced by the curved geometry \cite{tononi2023}. 
Indeed, the solution of one and two-body problems on a spherical surface evinced interesting consequences associated to curvature and to non-separability.
For instance, there were studies of $p$-wave dimers moving under a geometrically-induced gauge field \cite{shi2017}, of $s$-wave scattering of one body \cite{zhang2018} and of two-bodies on a large sphere \cite{tononi2022}, of the gas-to-soliton crossover \cite{tononi2024}, and of the anyonic spectrum on the sphere \cite{ouvry2019, polychronakos2020}. 
These developments address the fundamental theoretical issue of understanding few-body physics in curved geometries, and are potentially interesting for experiments with shell-shaped gases \cite{guo2020, carollo2022, jia2022, huang2025} and with other low-dimensional curved geometries \cite{fernholz2007}.

In this Letter, we calculate the 
energy and wave function of two atoms confined to a sphere varying the scattering length $a$ and the total angular momentum $j$. Technically, at fixed $j$ the problem is reduced to a finite set of coupled differential equations for the relative wave function. We derive these equations by adapting to our case the rigid-rotor formalism of Ref.~\cite{shi2017}. We find that for a small dimer, when $a$ is much smaller than $R/\sqrt{j}$ ($R$ is the sphere radius), the wave function is well approximated by the product of the isotropic relative wave function, the same as in the flat case, and the wave function of the center-of-mass motion with angular momentum $j$. 
However, upon increasing $a$ (or $j$) the dimer becomes anisotropic; the relative wave function gets more and more squeezed in the direction perpendicular to the direction of the center-of-mass motion. We argue that this squeezing is due to an effective harmonic confinement with oscillator length $\sim R/\sqrt{j}$ acting on the relative degree of freedom. For large $j$ our two-body problem on a sphere can be reduced to a flat-space quasi-one-dimensional problem in this effective harmonic confinement. 
The two-body state can be of localized or delocalized character and it can be two dimensional or quasi one dimensional depending on the relationships among the three relevant length scales, $a$, $R$, and $R/\sqrt{j}$. In the rest of the Letter we use the sphere radius as the unit of length, i.e., we set $R=1$.

\begin{figure}[hbtp]
\centering
\includegraphics[width=0.999\columnwidth]{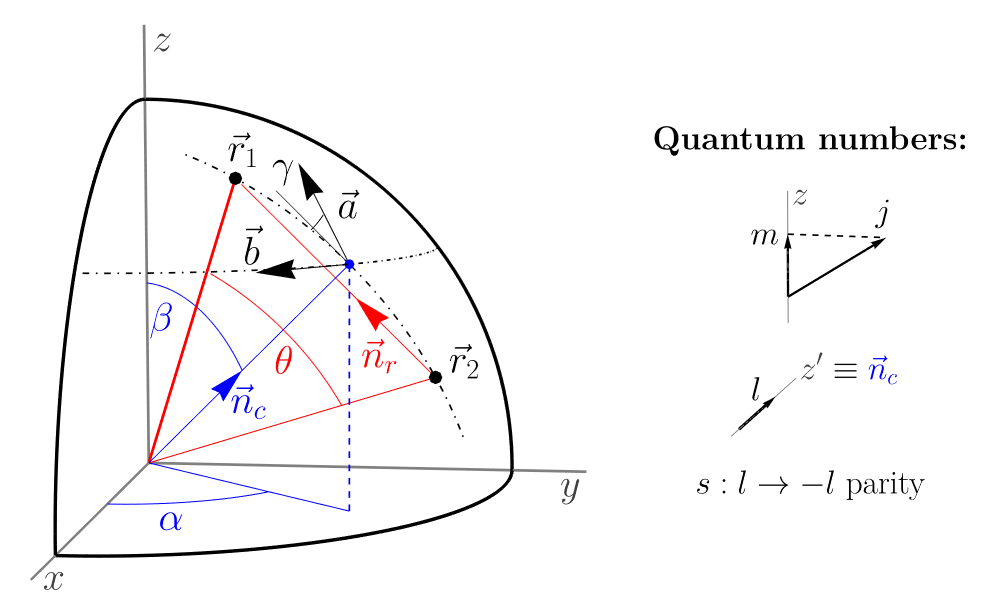}
\caption{Illustration of the coordinate system of two particles located at $\vec{r}_1$ and $\vec{r}_2$. Their geodesic center of mass, pointed by the vector $\vec{n}_c$ is described by the spherical coordinates $\alpha$ and $\beta$, while their relative position, pointed by the vector $\vec{n}_r$, is described by the angles $\gamma$ and $\theta$. 
The system is characterized by four quantum numbers: the total angular momentum $j$, its projection $m$ along the axis $z$, the angular momentum $l$ describing the molecular rotation along the $z' = \vec{n}_c$ axis, and the parity $s$ of the eigenstates under the exchange $l \to -l$.
}
\label{fig1}
\end{figure}

Our two-body system on a sphere has four angular degrees of freedom, which admit different parametrizations.  We first work with the center-of-mass and relative angles $\vec{u} = ( \alpha,\beta,\gamma,\theta )$. Figure~\ref{fig1} shows how the set $\vec{u}$ is related to the single-particle vectors $\vec{r}_1$ and $\vec{r}_2$ (see the Supplemental Material for explicit expressions). The spherical coordinates $\alpha \in [0,2\pi]$ and $\beta \in [0,\pi]$ parametrize the center-of-mass vector $\vec{n}_c = (\vec{r}_1+\vec{r}_2 ) /|\vec{r}_1+\vec{r}_2|$. 
The relative vector $\vec{n}_r = (\vec{r}_1-\vec{r}_2 )/|\vec{r}_1-\vec{r}_2|$ is parametrized by $\alpha$, $\beta$, and by the angle $\gamma \in [0,2\pi]$ between the geodesic passing by the center of mass and the great circle passing by the north pole. Finally, $\theta \in [0,\pi]$ is the relative angular distance between the atoms.

The Schr\"odinger equation for the two-body wave function $\Psi(\alpha,\beta,\gamma,\theta)$ reads
\begin{equation}
(\hat{T} - E)\Psi = 0,
\label{twobodyschrodinger}
\end{equation}
where the relation between the two-body energy $E$ and the $s$-wave scattering length $a$ is obtained by imposing the Bethe-Peierls boundary condition $\Psi |_{\theta \to 0} \propto \ln(\theta/a)$. 
The kinetic energy operator $\hat{T}$ is derived by directly calculating the Laplace-Beltrami operator in the coordinates $\vec{u}$, finding (see the Supplemental Material for details): 
$\hat{T} = (\hat{J}^2_{x'}/I_{x'} +\hat{J}^2_{y'}/I_{y'} +\hat{J}^2_{z'}/I_{z'})/2 + \hat{L}^2_{\theta}$,
which is the sum of the rotational energies along the molecular axes $( \vec{x}\,',\vec{y}\,',\vec{z}\,' ) = ( \vec{n}_r, \vec{n}_c \times \vec{n}_r,\vec{n}_c )$ and of the kinetic energy $\hat{L}^2_{\theta} = -(\sin\theta)^{-1} \partial_{\theta} (\sin\theta \, \partial_{\theta})$ for the relative motion along $\theta$~\cite{zhang2018}. 
The moments of inertia are equal to $I_{x'} = 2 \cos^2(\theta/2)$, $ I_{y'} = 2 $ and $I_{z'} = 2\sin^2(\theta/2)$, while the expressions of the angular momentum operators $\hat{J}_{x'}$, $\hat{J}_{y'}$ and $\hat{J}_{z'}$ in terms of $\alpha$, $\beta$ and $\gamma$ are reported in the Supplemental Material. We now rewrite $\hat{T}$ through the total angular momentum operator $\hat{J}^2 = \hat{J}_{x'}^2 + \hat{J}_{y'}^2 + \hat{J}_{z'}^2$ and the ladder operators $\hat{J}_{\pm} = \hat{J}_{x'} \pm i \hat{J}_{y'}$, obtaining
\begin{equation}
\hat{T} = A(\theta)\hat{J}^2 +B(\theta)\hat{J}^2_{z'} + C(\theta) (\hat{J}_{+}^2+\hat{J}_{-}^2) + \hat{L}^2_{\theta},
\label{T3}
\end{equation}
with $A(\theta) = [ 1/\cos^2(\theta/2) +1 ]/8$, $B(\theta) = [ 8/\sin^2\theta -1 - 3/\cos^2(\theta/2)]/8$ and $C(\theta) =  \tan^2(\theta/2) /16$. 
The common eigenstates of $\hat{J}^2$ and $\hat{J}_{z'}$ are the Wigner-D matrices $D_{ml}^{j*}(\alpha,\beta,\gamma)$ \cite{varshalovic1988}, satisfying the relations 
\begin{equation}\label{Properties}
   \begin{aligned} 
   \hat{J}^2 D_{ml}^{j*}& = j(j+1) D_{ml}^{j*},\\
   \hat{J}_{z'}^2 D_{ml}^{j*}&= l^2 D_{ml}^{j*},\\
   \hat{J}_{\pm} D_{ml}^{j*}&= [j(j+1) - l(l\pm 1)]^{1/2}D_{m l\pm 1}^{j*}.
   \end{aligned}
\end{equation}
These eigenstates are labeled by the total angular momentum $j$, by its projection along the $z$ axis $m=-j,...,j$, and by the angular momentum projection along the $z'$ axis $l=-j,...,j$ (see Fig.~\ref{fig1}). 
Note that the operator $\hat{T}$ conserves $j$ and $m$, but it does couple states with $l$ different by $2$. 
We decompose the wave function in each $j,m$ channel as \cite{Landau,cook}
\begin{widetext}
\begin{equation}\label{mEqj}
\Psi_{j,m}(\alpha,\beta,\gamma,\theta) = \sum_{l=0, \, \atop {l \text{ even}}}^{j} \psi_{l}(\theta) S_{jml}(\alpha,\beta,\gamma),
\end{equation}
where $S_{jml} = (D_{ml}^{j*} + D_{m-l}^{j*} ) / \sqrt{2}$ for $l>0$, while $S_{jm0} = D_{m0}^{j*}$, and using the properties (\ref{Properties}) we reduce Eq.~(\ref{twobodyschrodinger}) to 
\begin{equation}\label{Schr}
[\hat{L}^2_{\theta}+j(j+1)A(\theta) +l^2B(\theta)]\psi_l(\theta) + C(\theta)[c_l\psi_{l+2}(\theta)+c_{l-2}\psi_{l-2}(\theta)]=E_j \psi_l(\theta),
\end{equation}
\end{widetext}
where 
$c_l=\sqrt{(j-l-1) (j - l) (j + l+1) (j + l+2)}$, 
$c_0= \sqrt{2(j-1)j (j +1) (j +2)}$,
and $c_l = 0$ for $l<0$. 
Note that we only include the even-$l$ wave function components in Eq.~\eqref{mEqj} because the operator $\hat{T}$ does not couple even-$l$ channels to odd-$l$ channels.
Indeed, for zero-range $s$-wave interaction the odd-$l$ part describes noninteracting states  and we are only interested in the even-$l$ channels (the $p$-wave-interacting case has been considered in Ref.~\cite{shi2017}). 
In fact, the $s$-wave interaction is effective only in the equation with $l=0$ because the other components experience the centrifugal barrier $l^2B(\theta)\propto 1/\theta^2$. 
Also note that $\hat{T}$ conserves parity under the exchange $l\rightarrow -l$.
While the odd-parity configurations are insensitive to the interaction, the symmetric states under this exchange feel the interaction through their coupling to $\psi_0$. 
By expanding the wave function in terms of symmetric superpositions of opposite $l$ channels, $S_{jlm}$, we select only the even-parity configurations.
Thus, our dimer problem with $s$-wave interaction essentially reduces to $(j+2)/2$ coupled differential equations for even $j$ or to $(j+1)/2$ equations for odd $j$ (recall that $|l|\leq j$). 

In particular, for $j=0$ we have the single equation $(\hat{L}_{\theta}^2 - E_0) \psi_0(\theta) = 0$, solved in terms of Legendre functions~\cite{tononi2024, fedotova2001}: $\psi_0(\theta)\propto P_{-1/2+s}[\cos(\pi-\theta)]$, with $s=(E_0+1/4)^{1/2}$. The energy $E_0$ is then obtained from $\ln(1/a) = [\mathcal{D}(1/2+s) + \mathcal{D}(1/2-s)]/2 +\ln(e^{\gamma_E}/2)$, where $\mathcal{D}$ is the digamma function and $\gamma_E$ is the Euler-Mascheroni constant. 

The case $j=1$ is governed by a different but also single equation $[\hat{L}_{\theta}^2 +2A(\theta) - E_1] \psi_0(\theta) = 0${, which can be rewritten in the form of the Jacobi differential equation. 
We obtain its solution in terms of Jacobi functions $\psi_0(\theta) \propto (1+\cos\theta)^{-1/2} J_{\nu}^{(-1,0)}[\cos(\pi-\theta)]$, with $\nu=E_1^{1/2}$ \cite{referee}.
The Bethe-Peierls boundary condition leads to the relation between the energy and the scattering length: $\ln(1/a) = [\mathcal{D}(\nu) + \mathcal{D}(-\nu)]/2 +\ln(e^{\gamma_E}/2)$.}

\begin{figure}[hbtp]
\centering
\includegraphics[width=1.0\columnwidth]{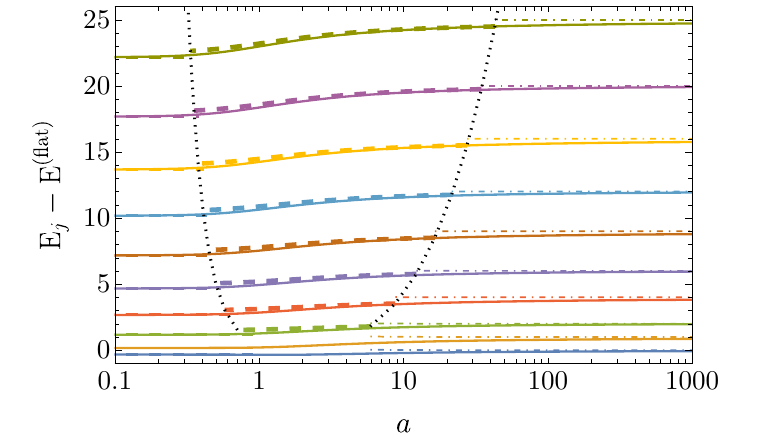}
\caption{Dimer energy spectrum versus $a$ for $j=0,...,9$ (continuous lines). 
The exact curves converge towards the analytical asymptotes of Eqs.~\eqref{energysmalla} and \eqref{energylargea} valid, respectively, in the strongly-attractive regime $a \ll 1$ (dashed lines), and in the noninteracting regime $a \gg 1$ (dot-dashed lines). 
In the intermediate $a$ regime, the curves are well reproduced by the semianalytical quasi-one-dimensional theory (thick dashed lines). The dimer is quasi-one-dimensional between the dotted curves; the left border corresponds to the dimer aspect ratio $\approx 1$ and the right border indicates where the dimer size becomes comparable to the sphere radius (see text). 
}
\label{fig2}
\end{figure}

For $j>1$ we solve Eqs.~(\ref{Schr}) numerically. The energies $E_j$ as functions of $a$ are presented in Fig.~\ref{fig2} as solid curves. The dashed curves correspond to the two leading-order terms in the expansion of the energy in powers of small $a$ 
\begin{equation}
E_{j}^{(a \ll 1)} = E^{(\text{flat})}-1/3 + j(j+1)/4,
\label{energysmalla}
\end{equation}
where $E^{(\text{flat})} = -4 \exp(-2 \gamma_E)/a^2$ is the dimer energy in the flat case. The center-of-mass energy $j(j+1)/4$ and the leading-order curvature-induced shift $-1/3$ can be obtained by solving Eq.~(\ref{Schr}) perturbatively at small $\theta\sim a$. In doing this it is convenient to rewrite the operator $\hat{L}_\theta^2$ and the functions $A(\theta)$, $B(\theta)$, $C(\theta)$ changing the variable from the angle $\theta$ to the chord distance $\rho = 2\sin(\theta/2)$ (see Ref.~\cite{tononi2024}). The dash-dotted horizontal lines correspond to the formulas
\begin{align}
\begin{split}
E_{j \, \text{even}}^{(a \gg 1)} &= j^2/4 + j/2, \qquad 
\\
E_{j \, \text{odd}}^{(a \gg 1)} &=  j^2/4 + j/2 + 1/4.
\label{energylargea}
\end{split}
\end{align}
Equations~(\ref{energylargea}) follow from the fact that the energy on the sphere scales quadratically with the angular momentum and, therefore, for fixed total angular momentum $j$ the lowest-energy state of two noninteracting atoms is obtained when their (integer) angular momenta $j_1$ and $j_2$ are as close as possible to $j/2$ and are also such that $j_1+j_2=j$.
In Fig.~\ref{fig2} we show the lowest-energy two-body states for fixed total angular momenta $j$. These energies do not depend on the projection, which can be $m=-j,...,j$. Note that we assume the thin-shell regime completely neglecting the degree of freedom perpendicular to the sphere surface (see Ref. \cite{carmesin2024} for an analysis of the case where the radial excitations are not frozen and where angular and radial degrees of freedom are coupled).

\begin{figure}[hbtp]
\centering
\includegraphics[width=1.0\columnwidth]{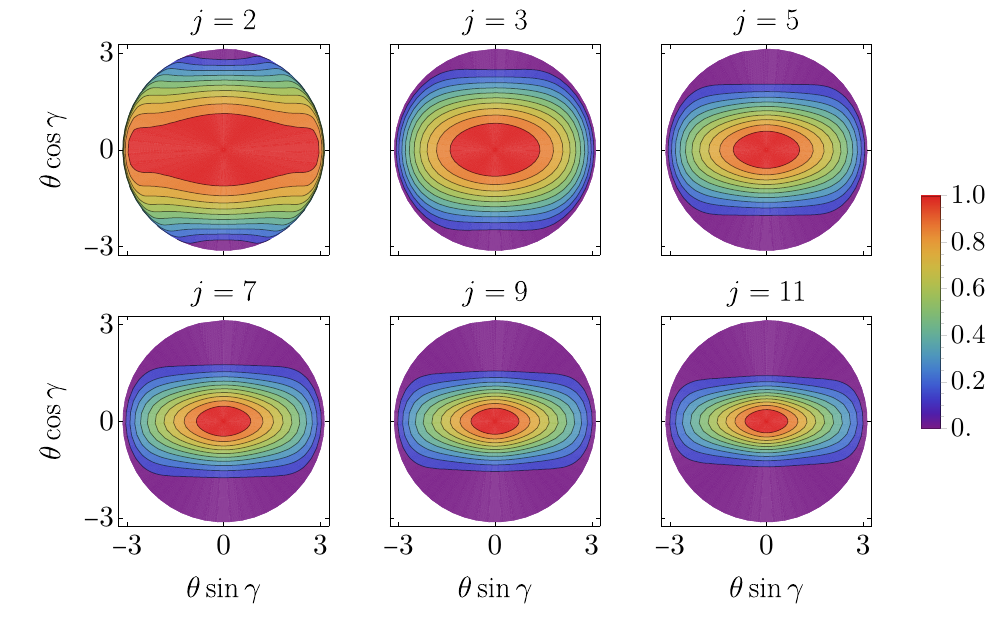}
\caption{
Contour plots of the ratio (rescaled to its maximum) $| \Psi_{j,m}(\alpha,\pi/2, \gamma, \theta) / K_0(2 e^{-\gamma_E}\theta/a) |$ for $a = 2$, which demonstrate the squeezing of the state along the direction of motion of the center of mass.
}
\label{fig3}
\end{figure}

We now discuss how the wave function depends on $j$. For $j=0$ and $j=1$ only $\psi_0$ is nonzero and the total wave function is independent of $\gamma$, which can be seen from Eq.~(\ref{mEqj}) bearing in mind that $S_{000} = 1$ and $S_{110} = e^{-i \alpha} \sin\beta /\sqrt{2}$. The dimer in these cases is isotropic although the $\theta$ dependence of its wave function is sensitive to $j$. The anisotropy first appears in the case $j=2$ where $\psi_2\neq 0$. It manifests itself in a squeezing of the molecule along a direction which depends on the center-of-mass angles $\alpha$ and $\beta$ and on $m$ (note, however, that $\psi_l$ depend on $j$, but not on $m$). The phenomenon can be seen clearly in the case $m=j$, which corresponds to the center-of-mass motion of the molecule along the equator. If we also set the center of mass on the equator ($\beta=\pi/2$), the wave function (\ref{mEqj}) explicitly reads $\Psi_{j,j}(\alpha,\pi/2,\gamma,\theta) \propto e^{-ij\alpha} [ \psi_{0}(\theta) / j! + \sum_{l>0} \psi_{l}(\theta)\cos(l\gamma) / \sqrt{2(j+l)!(j-l)!} ]$. We demonstrate the squeezing of the dimer by plotting the quantity $| \Psi_{j,j}(\alpha, \pi/2, \gamma, \theta) / \Psi^{\rm (flat)}(\theta)|$ in Fig.~\ref{fig3} for $a = 2$ (note that $|\Psi_{j,j}(\alpha,\pi/2, \gamma, \theta)|$ is independent of $\alpha$). 
We divide by the (isotropic) bound-state wave function in the flat-case limit $\Psi^{\rm (flat)}(\theta)=K_0(2e^{-\gamma_E}\theta/a)$ to remove the logarithmic divergence at $\theta \to 0$ and to better visualize the angular distribution of the state. We observe that by increasing $j$ the dimer becomes more and more squeezed in the direction perpendicular to the equator, i.e., perpendicular to the center-of-mass motion.

The squeezing becomes more pronounced for large $j$. In this case the dimer wave function (\ref{mEqj}) involves many $l$-components and for describing the system it is more convenient to switch from $(\alpha,\beta,\gamma,\theta)$ to the single-particle bases of polar and azimuthal angles $(\theta_1,\phi_1,\theta_2,\phi_2)$. As we have already mentioned, for fixed $j$ (let us assume for simplicity that $j$ is even), two non-interacting atoms prefer to occupy single-particle orbitals with angular momenta $j_1=j_2=j/2$. If $m=j$, we also have $m_1=m_2=j/2$. For large $j$ such orbitals are localized close to the equator of the sphere where $\theta_\sigma\approx \pi/2$. The variation of $\theta_\sigma$ is of order $|\theta_\sigma-\pi/2|\sim 1/\sqrt{j}$. One can see this by switching to the variable $y_\sigma=\theta_\sigma-\pi/2$ and observing that the single-particle kinetic-energy operator can be written as $[-\partial^2/\partial y_\sigma^2+\tan y_\sigma \partial/\partial y_\sigma +j^2/(4\cos^2y_\sigma)]/2$. The expansion of $j^2/8\cos^2y_\sigma$ at small $y_\sigma$ gives an approximately harmonic potential with frequency $j/2$ which localizes the wave function to the oscillator length $\sqrt{2/j}$. As we show in the Supplemental Material this localization persists also in the interacting case. The dimer problem becomes quasi-one-dimensional and the dimer energy $E=j^2/4+j/2+q^2$ is obtained by solving 
\begin{equation}
\label{Quasi1DEq}  
\int_0^\infty e^{2q^2\tau/j}\left(\sqrt{\tau/\sinh\tau}e^{\tau/2}-1\right)\frac{d\tau}{4\pi\tau}=\frac{1}{2\pi}\ln\frac{\sqrt{-q^2}ae^{\gamma_E}}{2}.
\end{equation}
The corresponding results are shown as dashed curves in Fig.~\ref{fig2}. Equation~(\ref{Quasi1DEq}) is valid for $j=m\gg 1$ and we require $1\ll -q^2\lesssim j$ (see more details in Appendix C). Under these conditions the two-body wave function is well approximated by $\Psi(\theta_1,\phi_1,\theta_2,\phi_2)\approx \exp[-j(\theta_1-\pi/2)^2/4-j(\theta_2-\pi/2)^2/4-\sqrt{-q^2}|\phi_1-\phi_2|]$, its quasi-one-dimensional character is explicit; in the direction of the center-of-mass motion the dimer has the size $1/\sqrt{-q^2}$ which is larger than its width given by $\approx 1/\sqrt{j}$.

We can now summarize the main regimes of an $s$-wave-interacting dimer on a sphere. For small $j$, with increasing $a$ the dimer increases in size, but remains to a large extent isotropic. The change of the character in this case happens at $a\sim 1$ when the dimer size becomes comparable to the sphere radius. For large $j$ we identify the following three regimes. For $a\lesssim 1/\sqrt{j}$ the dimer is strongly bound and approximately isotropic. In the interval $1/\sqrt{j} \lesssim a \lesssim a^*$ the dimer is quasi-one-dimensional and its size is smaller than the sphere radius. The characteristic scattering length $a^*\approx e^{\sqrt{\pi j/2}}$ is obtained by setting $q^2= -1$ in Eq.~(\ref{Quasi1DEq}). It marks the crossover to the third regime where the two atoms are delocalized along the equator, but localized in the perpendicular direction with polar angles $|\theta_\sigma-\pi/2|\sim 1/\sqrt{j}$. The dotted curves in Fig.~\ref{fig2} correspond to $a=1/\sqrt{j}$ (left border) and $a=a^*$ (right border) and indicate the regime where the dimer is quasi-one-dimensional. 

In conclusion, we find the spectrum and wave functions of an $s$-wave-interacting dimer on a spherical surface as a function of the scattering length $a$ and total angular momentum $j$. The nonseparability of the relative and center-of-mass degrees of freedom manifests itself in squeezing of the dimer in the direction transversal to the center-of-mass motion. The effect is most pronounced for $a\gtrsim 1/\sqrt{j}$ when the dimer becomes quasi-one-dimensional. Moreover, for $a\gg 1$, when the attraction is insufficient to localize two atoms into a dimer at low $j$, this transversal squeezing enhances the attraction and eventually leads to a bound quasi-one-dimensional dimer at sufficiently large $j$.

Our findings have implications for ongoing experiments with shell-shaped magnetic~\cite{guo2020, carollo2022, beregi2024} and optical~\cite{jia2022, huang2025} traps as well as for proposals based on quantum effects in self-bound mixtures~\cite{ma2025}. One can be able to create rapidly-rotating gases by combining gravity-compensation mechanisms with phase-imprinting techniques for transferring angular momentum to the gas \cite{sharma2024, debussy2024}. The two-body spectrum that we calculate can be measured experimentally by radio-frequency spectroscopy \cite{stoferle2006, thalhammer2006} and the anisotropy of the dimers can be observed in time-of-flight experiments. If we neglect interactions, after a long free expansion, the distribution of atoms is directly related to the Fourier transform of the original wavefunction. The free expansion of shell-shaped gases in their ground states at zero angular momentum has been analyzed in Refs.~\cite{lannert2007, tononi2020, carollo2022, jia2022}. In the future, to account for trap imperfections (gravitational sag, ellipticity, local variations of the curvature, finite shell thickness and its variations, etc.) we think of generalizing our harmonic large-$j$ theory to a finite out-of-surface confinement and to generic nonspherical geometries. 
From the many-body perspective, we can mention the study of the BCS-BEC crossover on a sphere \cite{he2022}, which would be interesting to reconsider at a finite angular momentum, and the gas-to-soliton transition for attractive bosons \cite{tononi2024}. 
This transition is characterized by a subtle interplay among the space curvature, mean-field and beyond-mean-field effects, and we believe that it should also be sensitive to the total angular momentum.

\begin{acknowledgments}
A.T. acknowledges financial support of the Horizon Europe programme HORIZON-CL4-2022-QUANTUM-02-SGA via the project 101113690 (PASQuanS2.1).
ICFO-QOT group acknowledges support from:
European Research Council AdG NOQIA; 
MCIN/AEI (PGC2018-0910.13039/501100011033,  CEX2019-000910-S/10.13039/501100011033, Plan National FIDEUA PID2019-106901GB-I00, Plan National STAMEENA PID2022-139099NB-I00, project funded by MCIN/AEI/10.13039/501100011033 and by the “European Union NextGenerationEU/PRTR" (PRTR-C17.I1), FPI); QUANTERA DYNAMITE PCI2022-132919, QuantERA II Programme co-funded by European Union’s Horizon 2020 program under Grant Agreement No 101017733;
Ministry for Digital Transformation and of Civil Service of the Spanish Government through the QUANTUM ENIA project call - Quantum Spain project, and by the European Union through the Recovery, Transformation and Resilience Plan - NextGenerationEU within the framework of the Digital Spain 2026 Agenda;
Fundació Cellex;
Fundació Mir-Puig; 
Generalitat de Catalunya (European Social Fund FEDER and CERCA program); 
Barcelona Supercomputing Center MareNostrum (FI-2023-3-0024); 
Funded by the European Union. Views and opinions expressed are however those of the author(s) only and do not necessarily reflect those of the European Union, European Commission, European Climate, Infrastructure and Environment Executive Agency (CINEA), or any other granting authority. Neither the European Union nor any granting authority can be held responsible for them (HORIZON-CL4-2022-QUANTUM-02-SGA PASQuanS2.1, 101113690, EU Horizon 2020 FET-OPEN OPTOlogic, Grant No 899794, QU-ATTO, 101168628). EU Horizon Europe Program (This project has received funding from the European Union’s Horizon Europe research and innovation program under grant agreement No 101080086 NeQSTGrant Agreement 101080086 — NeQST);
ICFO Internal “QuantumGaudi” project.
\end{acknowledgments}

\appendix

\renewcommand\thefigure{S\arabic{figure}}
\setcounter{figure}{0}
\renewcommand\theequation{S\arabic{equation}}
\setcounter{equation}{0}

\vspace{2mm}
\textbf{Supplemental Material: Dimer spectrum on a spherical surface}

\section{A) Particle positions in $\vec{u}$ coordinates}
\label{appA}
The particle positions $\vec{r}_1$ and $\vec{r}_2$ can be expressed in terms of $\vec{n}_c$ and $\vec{n}_r$ as
\begin{align}
\begin{split}
\vec{r}_1 &= \vec{n}_c \cos(\theta/2) + \vec{n}_r \sin(\theta/2),
\\
\vec{r}_2 &= \vec{n}_c \cos(\theta/2) - \vec{n}_r \sin(\theta/2),
\label{r1r2functionofu}
\end{split}
\end{align}
where the geodesic center-of-mass and relative vectors are respectively given by $\vec{n}_c = (\sin\beta \cos\alpha, \sin\beta \sin\alpha, \cos\beta )^T$ and $\vec{n}_r = \cos\gamma \, \vec{a} + \, \sin\gamma \, \vec{b}$. 
In particular, see Fig.~\ref{fig1}, $\vec{a} = (-\cos\beta \cos\alpha, -\cos\beta \sin\alpha, \sin\beta )^T$ is the tangent vector to the center of mass directed along the great circle passing by the north pole, while $\vec{b} = (\sin\alpha, -\cos\alpha, 0 )^T$ is the tangent vector to the center of mass directed along the circle parallel to the equator. 
Given the above relations, Eq.~\eqref{r1r2functionofu} represents the particles positions in terms of the angles $\vec{u}$. 

The body-fixed frame is built on the basis vectors $\vec{n}_r$, $\vec{n}_c \times \vec{n}_r$, and $\vec{n}_c$, which define the $x'$, $y'$, and $z'$ axes, respectively. The transition from the space-fixed to body-fixed frame is carried out with the help of the rotation matrix $\mathcal{R} = ( \vec{n}_r, \vec{n}_c \times \vec{n}_r,\vec{n}_c )$ such that any vector $\vec{r}\,' = (x', y', z')^T$ defined in the body-fixed frame corresponds to ${\vec{r}} = \mathcal{R} \, \vec{r}\,'$ in the laboratory frame. For instance, the particles coordinates correspond to $\vec{r_1}\,' = (\sin(\theta/2),0,\cos(\theta/2))^T$ and $\vec{r_2}\,' = (-\sin(\theta/2),0,\cos(\theta/2))^T$.

\section{B) Kinetic energy in $\vec{u}$ coordinates}
\label{appB}
The kinetic energy operator can be expressed in terms of the angles $\vec{u} = ( \alpha,\beta,\gamma,\theta )$ by calculating the Laplace-Beltrami operator
\begin{equation}
\hat{T} = -\frac{1}{2} \frac{1}{\sqrt{g}} \partial_i (\sqrt{g} g^{ij} \partial_j),
\label{LaplaceBeltramiOp}
\end{equation}
where $\partial_i = \partial/\partial u^i$, $g=\det(g_{ij})$, and $g^{ij}$ is the inverse of the metric tensor $g_{ij}$, defined through the line element squared $ds^2$ as $ds^2 = (d\vec{r}_1)^2 + (d\vec{r}_2)^2 = g_{ij} \, du^i du^j$.
Thus, by differentiating the coordinates at Eq.~\eqref{r1r2functionofu} in terms of the angles $\vec{u}$, we obtain the metric tensor
\begin{equation}
g_{ij} = \begin{bmatrix}
    h_{ij} & \vec{0} \\
    \vec{0} & 1/2
  \end{bmatrix},
\end{equation}
with $g =\sin^2\beta\sin^2\theta$, and where the symmetric $3\times 3$ tensor $h$ has components
\begin{align}
\begin{split}
h_{11} =& 2 \sin^2\beta[\cos^2\gamma+\cos^2(\theta/2)\sin^2\gamma-\sin^2(\theta/2)\cos(2\gamma)] +
\\
& 2\sin^2(\theta/2)\cos^2\beta,
\\
h_{12} =& 2 \sin^2(\theta/2) \sin\beta \sin\gamma \cos\gamma,
\\
h_{13} =& 2 \sin^2(\theta/2) \cos\beta,
\\
h_{22} =& 2 \sin^2\gamma+2\cos^2(\theta/2)\cos^2\gamma + 2\sin^2(\theta/2)\cos(2\gamma),
\\
h_{23} =& 0,
\\
h_{33} =& 2 \sin^2(\theta/2).
\end{split}
\end{align}
We calculate Eq.~\eqref{LaplaceBeltramiOp} explicitly and obtain the kinetic energy operator presented in the main text $\hat{T} = (\hat{J}^2_{x'}/I_{x'} +\hat{J}^2_{y'}/I_{y'} +\hat{J}^2_{z'}/I_{z'})/2 + \hat{L}^2_{\theta}$, whose angular momentum components are defined as 
\begin{align}
\begin{split}
\hat{J}_{x'} &= i \bigg( \frac{\cos\gamma}{\sin\beta}\partial_{\alpha} - \sin\gamma\partial_{\beta} - \cot\beta\cos\gamma\partial_{\gamma}  \bigg),
\\
\hat{J}_{y'} &= i \bigg( -\frac{\sin\gamma}{\sin\beta}\partial_{\alpha} - \cos\gamma\partial_{\beta} + \cot\beta\sin\gamma\partial_{\gamma} \bigg),
\\
\hat{J}_{z'} &= - i \partial_{\gamma},
\end{split}
\end{align}
in the molecular frame.

For completeness, we report the orthogonality relation of the Wigner-D functions used in the main text for projecting the Schr\"odinger equation \cite{varshalovic1988}
\begin{align}
\begin{split}
&\int_0^{2\pi} d\alpha \int_0^{\pi} d\beta \sin\beta \int_0^{2\pi} d\gamma  D_{m'l'}^{j'*}(\alpha,\beta,\gamma) D_{ml}^{j}(\alpha,\beta,\gamma) \\ &= \frac{8\pi^2}{2j+1} \delta_{jj'}\delta_{mm'}\delta_{ll'}.
\nonumber
\end{split}
\end{align}

\section{C) Derivation of Eq.~(\ref{Quasi1DEq})}
\label{appD}

In this appendix we discuss the case $j=m\gg 1$. Let us write the two-body wave function in the form $\Psi(\theta_1,\phi_1,\theta_2,\phi_2)=\chi(y_1,y_2,x)e^{ij\phi_c}$, where $y_\sigma=\theta_\sigma-\pi/2$ is the deviation from the equator, $x=\phi_1-\phi_2$, $\phi_c=(\phi_1+\phi_2)/2$, and $j$ is large integer, even or odd. The Schr\"odinger equation without interaction (\ref{twobodyschrodinger}) in these coordinates becomes
\begin{align}
\begin{split}
\label{SchrGen}
\sum_{\sigma=1,2}\left[-\frac{1}{2}\frac{\partial^2}{\partial y_\sigma^2}+\frac{\tan y_\sigma}{2} \frac{\partial}{\partial y_\sigma}+\frac{1}{\cos^2 y_\sigma}\left(\frac{j^2}{8}-\frac{1}{2}\frac{\partial^2}{\partial x^2}\right)\right]\chi
\\
-\frac{j}{2}\left(\frac{1}{\cos^{2}y_1} -\frac{1}{\cos^{2}y_2}\right)i\frac{\partial}{\partial x}\chi=E\chi.
\end{split}
\end{align}
The interaction is taken into account via a Bethe-Peierls boundary condition at $\{x,y_1-y_2\}={\bf 0}$. Let us assume (and a posteriori verify) that $y_\sigma\sim 1/\sqrt{j}$, $\partial/\partial y_\sigma\sim \sqrt{j}$, and that $\partial /\partial x$ is at most of order $\sqrt{j}$. Then, keeping only terms $\sim j^2\chi$ and $j\chi$ Eq.~(\ref{SchrGen}) reduces to 
\begin{equation}\label{SchrHarm}
\left(-\frac{1}{4}\frac{\partial^2}{\partial Y^2}+\frac{j^2}{4}Y^2-\frac{\partial^2}{\partial y^2}+\frac{j^2}{16}y^2-\frac{\partial^2}{\partial x^2}+\frac{j^2}{4}\right)\chi=E\chi
\end{equation}
with $y=y_1-y_2$ and $Y=(y_1+y_2)/2$. We thus arrive at the problem of two atoms of unit mass trapped in the $y$ direction by a harmonic potential with frequency $j/2$. As we mention in the main text this confinement arises from the expansion of the term $j^2/(8\cos^2 y_\sigma)$ in Eq.~(\ref{SchrGen}) in powers of $y_\sigma$. It reflects the centrifugal barrier felt by the atoms as they deviate from the equator trying to approach any of the poles. 

Equation~(\ref{SchrHarm}) is supplemented by the periodicity condition $\chi(Y,x,y)=(-1)^j \chi(Y,2\pi+x,y)$ and by the Bethe-Peierls constraint on the asymptotic behavior of the wave function $\chi(Y,x\rightarrow 0,y\rightarrow 0)\propto \ln[(x^2+y^2)/a^2]$. The center-of-mass motion separates from the relative one: $\chi(Y,x,y)=e^{-jY^2/2}\tilde\chi(x,y)$. The relative wave function $\tilde\chi$ can be written in the form of the Green function of a harmonic oscillator \cite{feynman1965} adapted to satisfy the periodicity condition 
\begin{equation}\label{GreenFunc}
\tilde\chi(x,y)=\sum_{n=-\infty}^\infty \!(-1)^{jn}\!\int_0^\infty \frac{e^{-y^2\frac{\coth\tau}{4l_\perp^2}+q^2l_\perp^2\tau+\frac{\tau}{2}-\frac{(x+2\pi n)^2}{4\tau l_\perp^2}}}{4\pi\sqrt{\tau\sinh\tau}} d\tau,
\end{equation}  
where $l_\perp=\sqrt{2/j}$ is the oscillator length. The total energy $E=j^2/4+j/2+q^2$ decomposes into the kinetic energy of the center-of-mass motion along the equator ($j^2/4$), the center-of-mass zero-point energy along $y$ ($j/4$), the relative zero-point energy ($j/4$), and the energy of the relative motion along $x$ which we denote by $q^2$.

We now establish the relation between $q^2$ and $a$ applying the Bethe-Peierls constraint, which is sufficient to write as $\tilde\chi(x,0)\propto \ln (x/a)$. Adding and subtracting the logarithmically diverging part from Eq.~(\ref{GreenFunc}) and then setting $x=0$ in the nondiverging terms gives
\begin{equation}\label{GreenFuncExp}
\tilde\chi(x,0)= F_1(x)+F_2+F_3+o(x^0),
\end{equation}
where
\begin{align}
\begin{split}
\label{Bessel}
F_1(x)&=\int_0^\infty e^{q^2l_\perp^2\tau-x^2/(4\tau l_\perp^2)}\frac{d\tau}{4\pi\tau}=K_0(\sqrt{-q^2l_\perp^2}x)/(2\pi)
\\
&= -\ln(\sqrt{-q^2x^2}e^{\gamma_E}/2)/(2\pi)+o(x^0),
\end{split}
\end{align}
\begin{equation}
\label{F_2}
F_2=\int_0^\infty e^{q^2l_\perp^2\tau}\left(\sqrt{\tau/\sinh\tau}e^{\tau/2}-1\right)\frac{d\tau}{4\pi\tau},
\end{equation}
and
\begin{align}
\begin{split}
\label{F_3}
F_3&=\sum_{n=-\infty, n\neq 0}^\infty \!(-1)^{jn}\!\int_0^\infty \frac{e^{q^2l_\perp^2\tau+\frac{\tau}{2}-\frac{(2\pi n)^2}{4\tau l_\perp^2}}}{4\pi\sqrt{\tau\sinh\tau}} d\tau
\\
&\approx \frac{1}{[e^{2\pi\sqrt{-q^2}}-(-1)^j]\sqrt{-2\pi q^2l_\perp^2}}.
\end{split}
\end{align}
In Eq.~(\ref{F_3}) we use the fact that the main contribution to the integral comes from $\tau\sim 1/l_\perp^2\gg 1$. Then, approximating $\sinh\tau \approx e^\tau/2$ the integral and the sum in Eq.~(\ref{F_3}) can be calculated analytically. The relation between $q$ and $a$ is obtained by noting that according to the Bethe-Peierls condition Eq.~(\ref{GreenFuncExp}) should behave as $-\ln(x/a)/(2\pi)$ at small $x$. In this manner we obtain
\begin{align}
\begin{split}
\label{FinalEq}  
\int_0^\infty e^{q^2l_\perp^2\tau}\left(\sqrt{\tau/\sinh\tau}e^{\tau/2}-1\right)\frac{d\tau}{4\pi\tau}
\\
+\frac{1}{[e^{2\pi\sqrt{-q^2}}-(-1)^j]\sqrt{-2\pi q^2l_\perp^2}}=\frac{1}{2\pi}\ln\frac{\sqrt{-q^2}ae^{\gamma_E}}{2}.
\end{split}
\end{align}

We now discuss validity of Eq.~(\ref{FinalEq}). When the distance between the two atoms is larger than $l_\perp$, i.e., when $|x+2\pi n|\gtrsim 1/\sqrt{j}$ (for any integer $n$), the wave function $\chi$ behaves as 
\begin{equation}\label{GreenFuncAsympt}
\chi(Y,x,y)\propto e^{-jY^2/2-jy^2/8}\sum_n (-1)^{nj}e^{-\sqrt{-q^2}|x+2\pi n|}.
\end{equation}
We derive Eq.~(\ref{GreenFuncAsympt}) from Eq.~(\ref{GreenFunc}) by using the approximations $\coth\tau \approx 1$ and $\sinh\tau \approx e^\tau/2$ valid since typical $\tau$ are large. We see that the characteristic length scale for the variation of $\chi$ in the $y$ direction is indeed $\sim 1/\sqrt{j}$ and the characteristic length scale in the $x$ direction is $1/|q|$. This verifies that Eq.~(\ref{SchrHarm}) is valid for $|q|\sim \partial/\partial x \lesssim \sqrt{j}$ as we initially assumed. 

We remind that passing from Eq.~(\ref{SchrGen}) to Eq.~(\ref{SchrHarm}) we kept only terms of order $j^2$ and $j$. Therefore, in principle, we should not allow $|q^2|$ to be smaller than $j$ in order not to exceed the accuracy of the approximation. { Under this condition the size of the dimer is smaller than the sphere radius and the exponentially small second term in the left-hand side of Eq.~(\ref{FinalEq}) can be neglected leading to Eq.~(\ref{Quasi1DEq}) of the main text, for which we require $-q^2\gg 1$.} However, considering the difference between Eqs.~(\ref{SchrHarm}) and (\ref{SchrGen}) as the perturbation and Eq.~(\ref{GreenFuncAsympt}) as the unperturbed solution to the harmonic problem Eq.~(\ref{SchrHarm}), one can show that the first-order and higher-order energy shifts are of order ${\rm max}\{q^2, 1\}/j$. We can thus claim that Eq.~(\ref{FinalEq}) also makes sense for $q^2\sim 1$ and can describe the whole crossover from the isotropic molecule ($|q|\sim \sqrt{j}$) to the noninteracting limit $a\rightarrow \infty$ where it correctly reproduces Eqs.~(\ref{energylargea}) predicting $q^2=0$ for even $j$ and $q^2=1/4$ for odd $j$. To solve Eq.~(\ref{FinalEq}) for positive $q^2$ we use analytical continuation.

\end{document}